# Fourier Neural Operator Networks: A Fast and General Solver for the Photoacoustic Wave Equation

Steven Guan, Ko-Tsung Hsu, and Parag V. Chitnis

S. Guan is with the Bioengineering Department, George Mason University., Fairfax, VA 22031 USA. and The MITRE Corporation., McLean, VA, 22102 (sguan2@gmu.edu)
K. Hsu is with the Bioengineering Department, George Mason University, Fairfax, VA 22030 USA. (khsu5@gmu.edu)
P. Chitnis is with the Bioengineering Department, George Mason University, Fairfax, VA 22030 USA. (pchitnis@gmu.edu)

## Abstract

Simulation tools for photoacoustic wave propagation have played a key role in advancing photoacoustic imaging by providing quantitative and qualitative insights into parameters affecting image quality. Classical methods for numerically solving the photoacoustic wave equation relies on a fine discretization of space and can become computationally expensive for large computational grids. In this work, we apply Fourier Neural Operator (FNO) networks as a fast data-driven deep learning method for solving the 2D photoacoustic wave equation in a homogeneous medium. Comparisons between the FNO network and pseudo-spectral time domain approach demonstrated that the FNO network generated comparable simulations with small errors and was several orders of magnitude faster. Moreover, the FNO network was generalizable and can generate simulations not observed in the training data.

## 1. Introduction

Photoacoustic imaging is a non-invasive hybrid imaging modality that combines the advantages of optical (e.g., high contrast and molecular specificity) and ultrasound (e.g., high penetration depth) imaging [1]. It has been applied for many preclinical and clinical imaging applications such as small-animal whole-body imaging, breast and prostate cancer imaging, and image-guided surgery [2]–[5]. Multispectral photoacoustic imaging can be used for functional imaging such as measuring blood-oxygen saturation and metabolism in biological tissues [6]. Photoacoustic imaging provides both structural and functional information that can potentially reveal novel insights into biological processes and disease pathologies [7].

In photoacoustic tomography (PAT), a tissue medium is illuminated using a short-pulsed laser. Optically absorbing molecules within the medium are excited and undergo thermoelastic expansion resulting in the generation of photoacoustic waves that are subsequently measured using an array of acoustic sensors [1]. An image representing the initial pressure distribution can be reconstructed from the measured time-dependent signals using analytical solutions, numerical methods, and model-based iterative methods [8]–[12]. A detailed understanding of parameters describing the imaging medium (e.g., optical, thermal, and acoustic properties of the tissue) and the imaging system (e.g., arrangement and characteristics of the laser source and acoustic sensors) is needed to reconstruct a high-quality PAT image.

PAT simulation is a highly useful tool that provides quantitative and qualitative insights into these parameters affecting image quality [13]. It is commonly used prior to experimentation and imaging to optimize the system configuration. It also plays an integral role in image reconstruction and provides numerical phantom data for the development of advanced algorithms such as iterative methods and deep learning methods [14]–[19]. Simulating the PAT image acquisition is comprised of two components, the optical illumination and photoacoustic propagation. For this work, we are primarily focused on the photoacoustic component. The equation for photoacoustic wave propagation can be solved numerically using classical methods such as the time domain finite element method [20], [21]. However, these methods can become computationally expensive, especially for large three-dimensional (3D) simulations.

Recently, deep learning has been explored as a computationally efficient partial differential equation (PDE) solver [22], [23]. It has the potential to revolutionize scientific disciplines and research by providing fast PDE solvers that approximate or enhance conventional ones. Applications requiring repeated evaluations of the forward model can greatly benefit from having reduced computation times. Here, we provide a brief overview of three deep learning methods for solving PDEs – finite dimensional operators, neural finite element models, and Fourier neural operators (FNO).

Finite dimensional operators use a deep convolutional neural network (CNN) to solve the PDE on a finite Euclidean Space [24], [25]. By definition, this approach is mesh-dependent, and the CNN needs to be retrained for solving the PDE at different resolutions and discretization. Neural finite element models are mesh-independent and closely resembles traditional finite element methods [22], [26]. It replaces the set of local basis functions in the finite element models with a fully connected neural network. It requires prior knowledge of the underlying PDE and is designed to solve for one specific instance of the PDE. The neural network needs to be retrained for new instances where the underlying PDE is parameterized with a different set of functional coefficients. FNO is a mesh-free approach that approximates the mapping between two infinite dimensional spaces from a finite collection of input-output paired observations [27], [28]. The neural operator is learned directly in the Fourier Space using a CNN. The same learned operator can be used without retraining to solve PDEs with different discretization and parameterization. Fourier Neural Operators have been demonstrated to achieve state-of-the-art results for a variety of PDEs (e.g., Burger's equation, Darcy Flow, and Navier-Stokes) and outperformed other existing deep learning methods [28].

To the best of our knowledge, this is the first paper that seeks to apply deep learning for solving the photoacoustic wave equation for simulating PAT. FNOs were chosen for this task given its flexibility in discretization and superior performance compared to other deep learning methods. Our contributions include adapting the FNO neural network and applying it as a fast PDE solver for simulating 2D photoacoustic wave propagation. Simulations from the FNO network and the widely used k-Wave toolbox for time domain acoustic wave propagation were compared in terms of accuracy and computation times. Further experiments were also conducted to evaluate the generalizability of the FNO network beyond the training data and the impact of key hyperparameters on network performance and complexity.

## 2. Methods

### A. Photoacoustic Signal Generation

The photoacoustic signal is generated by irradiating the medium with a nanosecond laser pulse. Chromophores excited by the laser undergo thermoelastic expansion and generate acoustic pressure waves. Assuming negligible thermal diffusion and volume expansion during illumination, the initial photoacoustic pressure $x$ can be defined as

$$x(r) = \Gamma(r)A(r) \quad (1)$$

where $A(r)$ is the spatial absorption function and $\Gamma(r)$ is the Grüneisen coefficient describing the conversion efficiency from heat to pressure [29]. The photoacoustic pressure wave $p(r,t)$ at position $r$ and time $t$ can be modeled as an initial value problem, where $c$ is the speed of sound [30].

$$(\partial_{tt} - c_0^2 \Delta)p(r,t) = 0, \quad p(r, t=0) = x, \quad \partial_t p(r, t=0) = 0 \quad (2)$$

Sensors located along a measurement surface, surrounding the medium, are used to measure a time-series signal. The linear operator $\mathcal{M}$ acts on $p(r,t)$ restricted to the boundary of the computational domain $\Omega$ over a finite time $T$ and provides a linear mapping to the measured time-dependent signal $y$.

$$y = \mathcal{M}_{p|\partial\Omega \times (0,T)} = Ax \quad (3)$$

### B. Conventional Methods for Solving the Wave Equation

Numerical approaches such as the finite-difference and finite-element methods are commonly used to solve PDEs by discretizing the space into a grid [31]. However, these methods are often slow for time domain modeling of broadband or high-frequency waves due to the need for a fine grid with small time-steps [13]. Computational efficiency can be improved using pseudo-spectral and k-space methods. The pseudo-spectral method fits a Fourier series to the data and reduces the number of grid points per wavelength required for an accurate solution [32]. The k-space method incorporates *a priori* information regarding the governing wave equation into the solution [33]. This allows for larger time steps and improves numerical stability in the case of acoustically heterogeneous mediums. The k-Wave toolbox, a widely used MATLAB tool for photoacoustic simulations, uses the pseudo-spectral k-space approach for solving time-domain photoacoustic wave simulations [34].

### C. Fourier Neural Operator Network

The FNO network was adapted for solving the 2D photoacoustic wave equation [28]. In our version, the FNO network does not apply Gaussian normalization to either the input or output of the training example. The network begins by mapping the input into a higher dimensional representation using a fully connected layer (Fig. 1). The transformed features are then iteratively updated by passing them through four successive Fourier layers. Finally, the updated features are projected to the desired dimensions using a fully connected layer. Through a combination of linear, Fourier, and non-linear transformations, the Fourier neural operators can approximate complex operators in PDEs that are highly non-linear with high frequency modes.

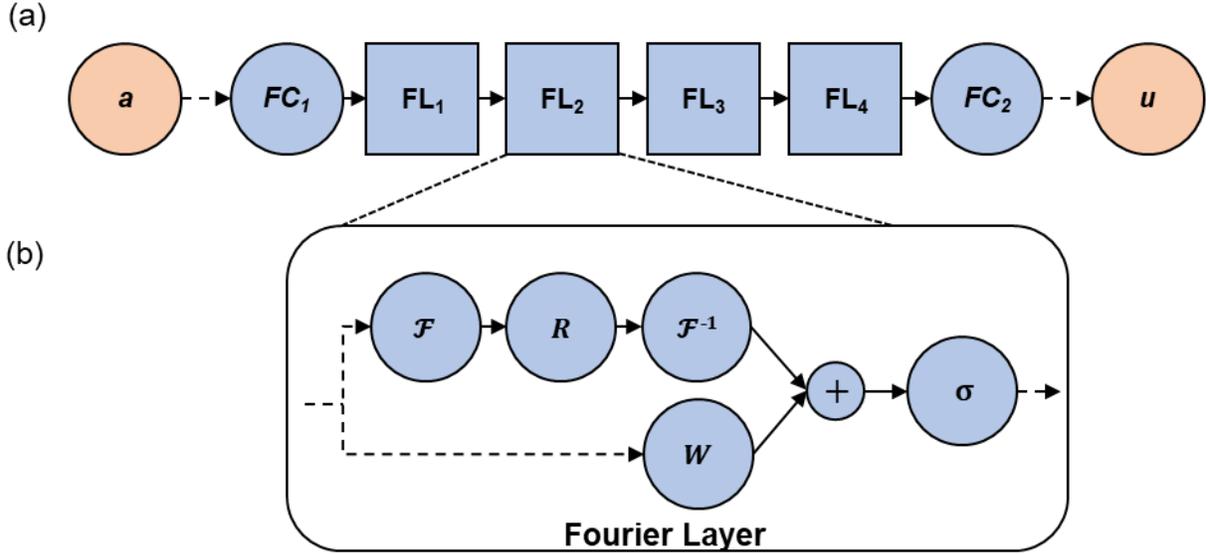

**Figure 1.** (a) Neural network architecture for the FNO network. The input *a* is mapped to a higher dimensional space using a fully connected layer ($FC_1$). The transformed feature is passed through four Fourier Layers (FL). Finally, a fully connected layer ($FC_2$) is used to obtain the final output *u* with the desired dimensions. (b) Architecture of a Fourier layer. The input goes through two paths in the Fourier layer. In the top path, the input undergoes a Fourier Transform $\mathcal{F}$, linear transform $R$, and inverse Fourier Transform $\mathcal{F}^{-1}$. In the bottom path, the input undergoes a linear transform $W$. Outputs from each path are summed together and undergo ReLU activation $\sigma$.

The photoacoustic wave equation can be solved with the FNO network using either a 2D or 3D implementation. In 2D, the FNO network performs 2D convolutions in space and finds a solution for some fixed interval length $\Delta t$. The solution is then recurrently propagated in time and used to solve for the next interval length. In 3D, the FNO network performs 3D convolutions in space-time and can directly output the full time series solution with any time discretization. While both implementations were demonstrated to have similar performance, the 3D FNO network was used in this work because it was found to be more expressive and easier to train [28]. The FNO network was implemented in Python v3.8 using the deep learning library PyTorch v1.7.1. The Adam optimizer with a mean squared error loss function was used to train the FNO network for 2,000 epochs over approximately two days on a NVIDIA Tesla K80 graphics processing unit (GPU).

Channels and modes are the two main hyperparameters that impact the accuracy of the FNO network. The channels parameter defines the width of the FNO network meaning the number of features learned in each layer. The modes parameter defines the number of lower

Fourier modes retained when truncating the Fourier series. The allowable maximum number of modes is related to the size of the simulation computational grid. In this work, the FNO network is assumed to have 64 modes and 5 channels unless otherwise specified.

### D. Photoacoustic Data for Training and Testing

The MATLAB toolbox k-Wave was used for photoacoustic wave simulation and to generate data for training and testing the FNO network [13]. The simulation medium was defined as a 64x64 computational grid, non-absorbing, and homogenous with a speed of sound of 1480 m/s and density of 1000 kg/m$^3$. Simulations were performed with a timestep of 20 ns for T=151 steps. The initial photoacoustic pressure was initialized using anatomically realistic breast vasculature phantoms that were numerically generated [35]. The training dataset (N=500) and testing dataset (N=100) were comprised of images representing the initial photoacoustic pressure (the input to the FNO network), and the corresponding simulation of the photoacoustic wave propagation (output of the FNO network). The FNO output was compared against the photoacoustic simulation performed using k-Wave, which served as the ground truth. While the training and testing data shared similar features, each example was unique. Simulations for the Shepp-Logan, synthetic vasculature, tumor, and Mason-M phantoms were also generated to evaluate the generalizability of the FNO network [13], [36].

## 3. Results

### A. Comparison of FNO Network and k-Wave Simulations

When tested on breast-vascular images similar to those used for training, the photoacoustic-wave simulations produced by the FNO network and k-Wave were remarkably similar and almost visually identical (Fig. 2). This demonstrated that the FNO network can model both broadband and high-frequency waves required for photoacoustic simulations. The FNO network and k-Wave simulations were quantitatively compared using the mean squared error (MSE). For the testing dataset, the MSE of the FNO network was 3.1e-5 which indicates that the FNO network was able to accurately simulate photoacoustic wave propagation.

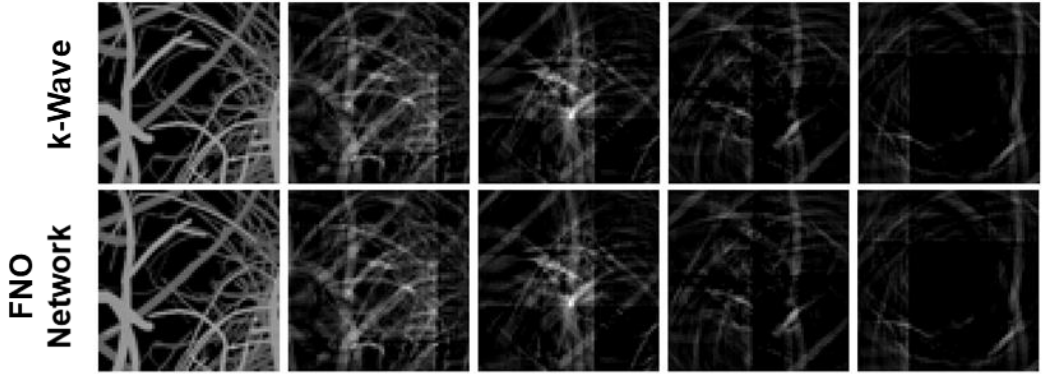

**Figure 2.** Visual comparison of the ground truth (Top Row) using k-Wave and the FNO network (Bottom Row) simulated photoacoustic wave propagation for an example vasculature image in a homogeneous medium at T = 1, 20, 40, 60, and 80 timesteps. The MSE for this example was 2.5e-5.

The time required to solve the photoacoustic wave equation largely depends on the discretization of the computational grid. In k-Wave, a simulation using a 64x64 grid required ~1.17 seconds to complete on a GPU. For a comparable simulation, the FNO network only required ~0.029 seconds to complete on a GPU which is approximately a 40x reduction in computation time.

B.  PAT Images Reconstructed from Simulations

For further validation, an *in-silico* experiment of PAT imaging with a 64-sensor linear array was conducted using the k-Wave and the FNO network simulations. Other sensor arrays and geometries can be used, but the linear array was chosen since it is widely available and used in laboratories. The sensor data for image reconstruction was created by sampling the photoacoustic pressures along the top row of the computational grid in each simulation. Images were then reconstructed from the time-series sensor data using the time reversal method in k-Wave [13]. The reconstructed images were highly similar with only minor differences (Fig. 3). The vasculature structures and limited-view artifact patterns seen in the FNO network image clearly matched those in the reconstructed image obtained using the k-Wave simulation data. The reconstructed images were quantitatively compared using MSE and the structural similarity index metric (SSIM), a metric ranging from 0 to 1 that measures the similarity between two images based on factors relevant to human visual perception (e.g., structure, contrast, and luminance) [37]. For the testing dataset (N=100), the FNO network images had a MSE of 3.1e-5 and SSIM of 0.99. This demonstrated that the time-series sensor data produced using the FNO network and k-Wave simulations were effectively identical.

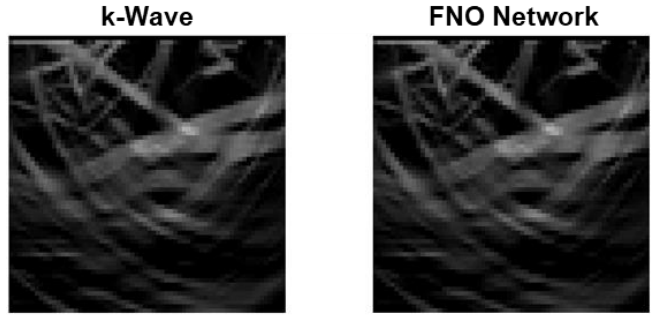

**Figure 3.** Images reconstructed using sampled sensor data from the k-Wave and FNO network photoacoustic simulations. The images were normalized to have intensities between 0 and 1. For this example, the MSE and SSIM were respectively 6.1e-5 and 0.99.

C. FNO Network Generalizability

    The FNO network was used to simulate photoacoustic wave propagation from Shepp-Logan, synthetic vasculature, tumor, and Mason-M phantoms. These phantoms contain many features not observed in the training dataset (breast vasculature). The FNO network and k-Wave simulations were visually similar for each phantom tested (Fig. 4). The MSE of the FNO network simulations were 2.1e-4 (Shepp-Logan), 3.7e-4 (synthetic vasculature), 1.9e-4 (tumor), and 6.4e-4 (Mason-M). These results provide evidence that the FNO network was generalizable to initial photoacoustic sources not in the training data.

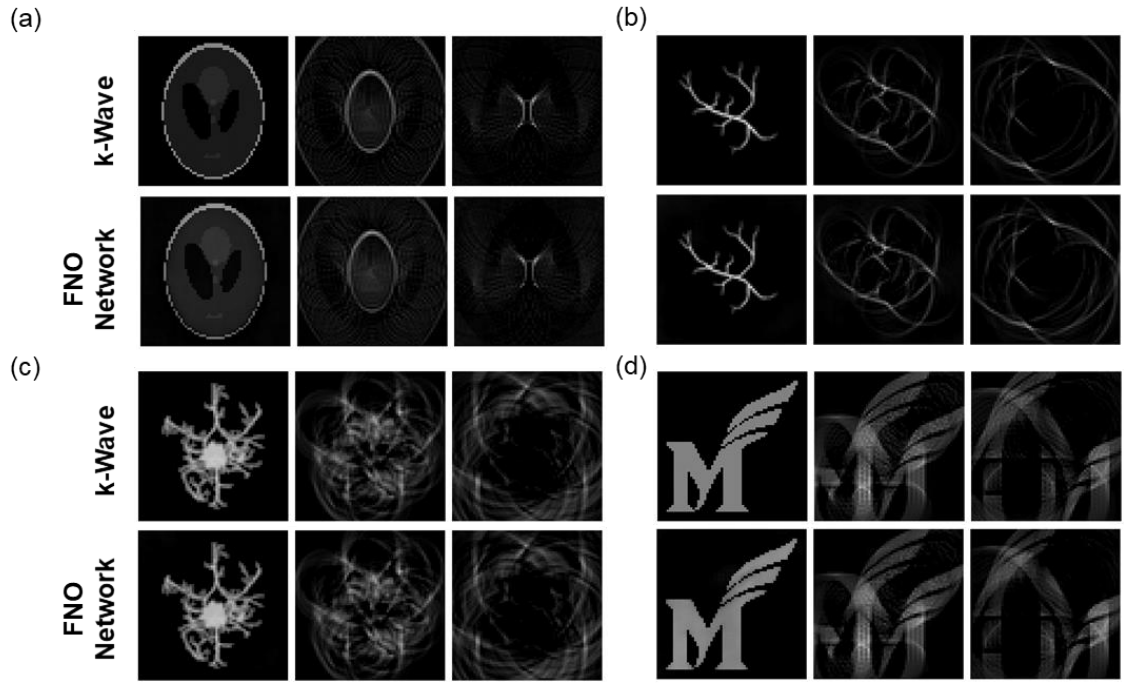

**Figure 4.** Comparison between FNO Network and k-Wave simulations for initial pressure sources using the (a) Shepp-Logan, (b) synthetic vasculature, (c) tumor, and (d) Mason-M phantoms at T=1,10, and 20 timesteps.

D. Hyperparameter Optimization

A study was conducted to investigate the impact of hyperparameter selection on the FNO network's accuracy. The number of modes had the largest impact since it is directly related to the truncation error in a Fourier layer. Networks with a lower mode produced simulations with a blurred appearance due to the loss of high frequency information (Fig. 5). Increasing the number of channels generally improved the FNO network's performance but also required more GPU memory (Table 1). There was no benefit in having an FNO network with more than five channels. Interestingly, the computation time to complete a simulation was approximately the same for FNO networks with a lower number of modes or channels. There was a moderate increase in computation time for the larger FNO networks with 64 modes and higher number of channels.

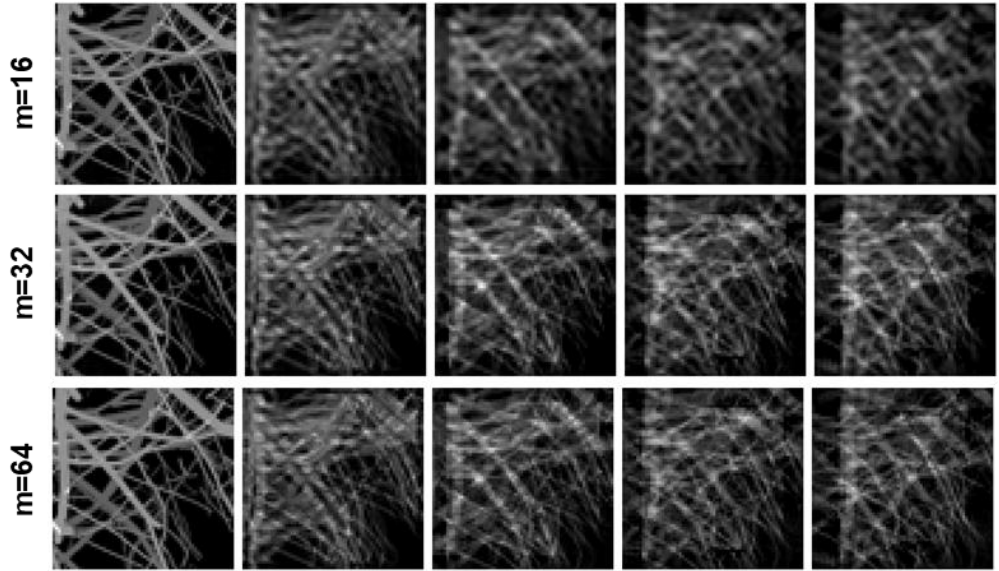

**Figure 5.** Visual comparison of photoacoustic wave simulations at T = 1, 5, 10, 15, and 20 timesteps. The FNO networks were parametrized with channels=5 and modes=16, 32, and 64.

**Table 1.** Comparison of FNO networks for different hyperparameters

| Modes | Channels | MSE | Time (s) | GPU Memory (GB) |
|---|---|---|---|---|
| 16 | 5 | 1.0e-3 | 0.022 | 1.3 |
| 32 | 5 | 1.1e-4 | 0.019 | 1.7 |
| 64 | 5 | 3.1e-5 | 0.022 | 4.8 |
| 64 | 2 | 2.7e-4 | 0.022 | 1.8 |
| 64 | 3 | 1.3e-4 | 0.021 | 2.5 |
| 64 | 4 | 4.4e-5 | 0.021 | 3.5 |
| 64 | 5 | 3.1e-5 | 0.022 | 4.8 |
| 64 | 6 | 3.5e-5 | 0.023 | 6.3 |
| 64 | 7 | 2.8e-5 | 0.026 | 8.2 |
| 64 | 8 | 3.6e-5 | 0.028 | 10.2 |

## 4. Discussions and Conclusions

Solving the 2D photoacoustic wave equation with traditional methods typically require a fine discretization of the computational grid and can be time-consuming to complete. Deep learning methods directly learn from data to solve PDEs and can be orders of magnitude faster with a minimal loss in accuracy. In this work, we applied the FNO network as a fast PDE solver for the 2D photoacoustic wave equation in a homogeneous medium. The FNO network and k-Wave solutions were qualitatively and quantitatively comparable. PAT images reconstructed from the FNO network and k-Wave simulations were effectively identical. This demonstrates that the sampled sensor data contained essentially the same information, and errors in the FNO network simulations did not impact the quality of images reconstructed. The FNO network was about 40x faster than k-Wave in completing a simulation with a 64x64 computational grid.

Applications requiring repeated evaluations of the photoacoustic wave equation such as iterative image reconstruction can be accelerated using the FNO network.

Model generalizability is a highly desirable property because it removes the need to retrain the model when it is used on examples not observed in the training data. This is important since the goal of the FNO network is to be like traditional methods as a general PDE solver for any arbitrary initial pressure source. The FNO network's generalizability was evaluated by having it perform photoacoustic simulations for four phantoms not in the training data (e.g., Shepp-Logan, synthetic vasculature, tumor, and Mason-M). The FNO network and k-Wave simulations were highly similar indicating that a trained FNO network can be used for simulations with any arbitrary initial pressure source. Furthermore, this provided evidence that the FNO network was learning the operator for photoacoustic wave propagation and not mainly specific solutions related to the training data.

Hyperparameter optimization is important to achieve the required level of accuracy and to minimize the memory required for training and inferring. The FNO network is parameterized by the number of modes and channels, and increasing either parameter typically improves model performance. In general, a higher number of modes is preferred but can be reduced if only a lower-resolution approximation of the solution is needed. Hyperparameter optimization is likely more important for simulations with large computational grids when limited GPU memory can be a problem. Alternative network architectures that are more memory efficient such as the recurrent 2D FNO network can be explored [28].

In this work, the FNO network was trained for solving the 2D acoustic wave equation in a homogeneous medium. Simulations with homogeneous mediums are widely used in many applications such as image reconstruction where the spatial distribution of heterogeneities is often unknown. Nevertheless, the FNO network can be used for simulations with heterogeneous mediums. The spatial distribution of heterogeneous medium properties can be provided as an input to the FNO network. By providing training examples of simulations with varying heterogeneous mediums, the FNO network likely can learn to solve the 2D wave equation and account for effects due to the heterogeneous medium.

A practical limitation in data-driven PDE solvers such as the FNO network is the need for high quality training data. Traditional solvers are often used to create arbitrarily large datasets to train the network. Depending on the size of the computational grid, this can be computationally formidable such as the case of 3D photoacoustic simulations. To create a large dataset in these scenarios, a high-performance computing environment would be needed to generate the training data in a reasonable timeframe.

## 5. Funding source

This research did not receive any specific grant from funding agencies in the public, commercial, or not-for-profit sectors.

## 6. Declaration of Competing Interest

The authors report no declarations of interest.

## 7. Acknowledgements

This project was supported by resources provided by the Office of Research Computing at George Mason University (URL: https://orc.gmu.edu). The authors would like to acknowledge Matthias Eyassu from the George Mason Biomedical Imaging Laboratory for providing the breast vasculature phantoms.

## 8. References

[1] J. Xia, J. Yao, and L. V. Wang, "Photoacoustic tomography: principles and advances," *Electromagn Waves (Camb)*, vol. 147, pp. 1–22, 2014.

[2] J. Xia and L. V. Wang, "Small-animal whole-body photoacoustic tomography: a review," *IEEE Trans Biomed Eng*, vol. 61, no. 5, pp. 1380–1389, May 2014, doi: 10.1109/TBME.2013.2283507.

[3] N. Nyayapathi and J. Xia, "Photoacoustic imaging of breast cancer: a mini review of system design and image features," *J Biomed Opt*, vol. 24, no. 12, Dec. 2019, doi: 10.1117/1.JBO.24.12.121911.

[4] B. L. Bungart *et al.*, "Photoacoustic tomography of intact human prostates and vascular texture analysis identify prostate cancer biopsy targets," *Photoacoustics*, vol. 11, pp. 46–55, Aug. 2018, doi: 10.1016/j.pacs.2018.07.006.

[5] C. Moore and J. V. Jokerst, "Strategies for Image-Guided Therapy, Surgery, and Drug Delivery Using Photoacoustic Imaging," *Theranostics*, vol. 9, no. 6, pp. 1550–1571, Feb. 2019, doi: 10.7150/thno.32362.

[6] M. Li, Y. Tang, and J. Yao, "Photoacoustic tomography of blood oxygenation: A mini review," *Photoacoustics*, vol. 10, pp. 65–73, Jun. 2018, doi: 10.1016/j.pacs.2018.05.001.

[7] L. V. Wang, "Prospects of photoacoustic tomography," *Med Phys*, vol. 35, no. 12, pp. 5758–5767, Dec. 2008, doi: 10.1118/1.3013698.

[8] S. Li, B. Montcel, W. Liu, and D. Vray, "Analytical model of optical fluence inside multiple cylindrical inhomogeneities embedded in an otherwise homogeneous turbid medium for quantitative photoacoustic imaging," *Opt Express*, vol. 22, no. 17, pp. 20500–20514, Aug. 2014, doi: 10.1364/OE.22.020500.

[9] M. Xu and L. V. Wang, "Universal back-projection algorithm for photoacoustic computed tomography," *Physical Review E*, vol. 71, no. 1, Jan. 2005, doi: 10.1103/PhysRevE.71.016706.

[10] Y. Hristova, P. Kuchment, and L. Nguyen, "Reconstruction and time reversal in thermoacoustic tomography in acoustically homogeneous and inhomogeneous media," *Inverse Problems*, vol. 24, no. 5, p. 055006, 2008, doi: 10.1088/0266-5611/24/5/055006.

[11] C. Huang, K. Wang, R. W. Schoonover, L. V. Wang, and M. A. Anastasio, "Joint Reconstruction of Absorbed Optical Energy Density and Sound Speed Distributions in Photoacoustic Computed Tomography: A Numerical Investigation," *IEEE Transactions on Computational Imaging*, vol. 2, no. 2, pp. 136–149, Jun. 2016, doi: 10.1109/TCI.2016.2523427.

[12] B. T. Cox and B. E. Treeby, "Artifact Trapping During Time Reversal Photoacoustic Imaging for Acoustically Heterogeneous Media," *IEEE Transactions on Medical Imaging*, vol. 29, no. 2, pp. 387–396, Feb. 2010, doi: 10.1109/TMI.2009.2032358.

[13] B. E. Treeby and B. T. Cox, "k-Wave: MATLAB toolbox for the simulation and reconstruction of photoacoustic wave fields," *J Biomed Opt*, vol. 15, no. 2, p. 021314, Apr. 2010, doi: 10.1117/1.3360308.

[14]   M. Xu, Y. Xu, and L. V. Wang, "Time-domain reconstruction algorithms and numerical simulations for thermoacoustic tomography in various geometries," *IEEE Transactions on Biomedical Engineering*, vol. 50, no. 9, pp. 1086–1099, Sep. 2003, doi: 10.1109/TBME.2003.816081.

[15]   G. Paltauf, J. A. Viator, S. A. Prahl, and S. L. Jacques, "Iterative reconstruction algorithm for optoacoustic imaging," *The Journal of the Acoustical Society of America*, vol. 112, no. 4, pp. 1536–1544, Sep. 2002, doi: 10.1121/1.1501898.

[16]   S. Guan, A. Khan, S. Sikdar, and P. Chitnis, "Fully Dense UNet for 2D Sparse Photoacoustic Tomography Artifact Removal," *IEEE J Biomed Health Inform*, Apr. 2019, doi: 10.1109/JBHI.2019.2912935.

[17]   S. Guan, A. A. Khan, S. Sikdar, and P. V. Chitnis, "Limited-View and Sparse Photoacoustic Tomography for Neuroimaging with Deep Learning," *Scientific Reports*, vol. 10, no. 1, Art. no. 1, May 2020, doi: 10.1038/s41598-020-65235-2.

[18]   S. Antholzer, M. Haltmeier, and J. Schwab, "Deep learning for photoacoustic tomography from sparse data," *Inverse Problems in Science and Engineering*, vol. 27, no. 7, pp. 987–1005, Jul. 2019, doi: 10.1080/17415977.2018.1518444.

[19]   A. Hauptmann *et al.*, "Model-Based Learning for Accelerated, Limited-View 3-D Photoacoustic Tomography," *IEEE Transactions on Medical Imaging*, vol. 37, pp. 1382–1393, 2018, doi: 10.1109/TMI.2018.2820382.

[20]   B. Baumann, M. Wolff, B. Kost, and H. Groninga, "Finite element calculation of photoacoustic signals," *Appl. Opt., AO*, vol. 46, no. 7, pp. 1120–1125, Mar. 2007, doi: 10.1364/AO.46.001120.

[21]   W. Xia *et al.*, "An optimized ultrasound detector for photoacoustic breast tomography," *Med Phys*, vol. 40, no. 3, p. 032901, Mar. 2013, doi: 10.1118/1.4792462.

[22]   M. Raissi, P. Perdikaris, and G. E. Karniadakis, "Physics-informed neural networks: A deep learning framework for solving forward and inverse problems involving nonlinear partial differential equations," *Journal of Computational Physics*, vol. 378, pp. 686–707, Feb. 2019, doi: 10.1016/j.jcp.2018.10.045.

[23]   D. Greenfeld, M. Galun, R. Basri, I. Yavneh, and R. Kimmel, "Learning to Optimize Multigrid PDE Solvers," in *International Conference on Machine Learning*, May 2019, pp. 2415–2423. Accessed: May 22, 2021. [Online]. Available: http://proceedings.mlr.press/v97/greenfeld19a.html

[24]   Y. Khoo, J. Lu, and L. Ying, "Solving parametric PDE problems with artificial neural networks," *Eur. J. Appl. Math*, vol. 32, no. 3, pp. 421–435, Jun. 2021, doi: 10.1017/S0956792520000182.

[25]   J. Adler and O. Öktem, "Solving ill-posed inverse problems using iterative deep neural networks," *Inverse Problems*, vol. 33, no. 12, p. 124007, 2017, doi: 10.1088/1361-6420/aa9581.

[26]   W. E and B. Yu, "The Deep Ritz method: A deep learning-based numerical algorithm for solving variational problems," *arXiv:1710.00211 [cs, stat]*, Sep. 2017, Accessed: May 22, 2021. [Online]. Available: http://arxiv.org/abs/1710.00211

[27]   L. Lu, P. Jin, and G. E. Karniadakis, "DeepONet: Learning nonlinear operators for identifying differential equations based on the universal approximation theorem of operators," *arXiv:1910.03193 [cs, stat]*, Apr. 2020, Accessed: May 22, 2021. [Online]. Available: http://arxiv.org/abs/1910.03193

[28]   Z. Li *et al.*, "Fourier Neural Operator for Parametric Partial Differential Equations," *arXiv:2010.08895 [cs, math]*, Oct. 2020, Accessed: Dec. 29, 2020. [Online]. Available: http://arxiv.org/abs/2010.08895

[29]   P. Beard, "Biomedical photoacoustic imaging," *Interface Focus*, vol. 1, no. 4, pp. 602–631, Aug. 2011, doi: 10.1098/rsfs.2011.0028.

[30]   M. Xu and L. V. Wang, "Universal back-projection algorithm for photoacoustic computed tomography," Apr. 2005, vol. 5697, pp. 251–255. doi: 10.1117/12.589146.

[31]   E. Tadmor, "A review of numerical methods for nonlinear partial differential equations," *Bull. Amer. Math. Soc.*, vol. 49, no. 4, pp. 507–554, 2012, doi: 10.1090/S0273-0979-2012-01379-4.

[32]   B. E. Treeby and J. Pan, "A practical examination of the errors arising in the direct collocation boundary element method for acoustic scattering," *Engineering Analysis with Boundary Elements*, vol. 33, no. 11, pp. 1302–1315, Nov. 2009, doi: 10.1016/j.enganabound.2009.06.005.

[33]   T. D. Mast, L. P. Souriau, D.-L. D. Liu, M. Tabei, A. I. Nachman, and R. C. Waag, "A k-space method for large-scale models of wave propagation in tissue," *IEEE Transactions on Ultrasonics, Ferroelectrics, and Frequency Control*, vol. 48, no. 2, pp. 341–354, Mar. 2001, doi: 10.1109/58.911717.

[34]   B. E. Treeby, J. Jaros, A. P. Rendell, and B. T. Cox, "Modeling nonlinear ultrasound propagation in heterogeneous media with power law absorption using a k-space pseudospectral method," *J Acoust Soc Am*, vol. 131, no. 6, pp. 4324–4336, Jun. 2012, doi: 10.1121/1.4712021.

[35]   A. Badano *et al.*, "Evaluation of Digital Breast Tomosynthesis as Replacement of Full-Field Digital Mammography Using an In Silico Imaging Trial," *JAMA Netw Open*, vol. 1, no. 7, p. e185474, Nov. 2018, doi: 10.1001/jamanetworkopen.2018.5474.

[36]   S. Arridge *et al.*, "Accelerated high-resolution photoacoustic tomography via compressed sensing," *Phys. Med. Biol.*, vol. 61, no. 24, p. 8908, 2016, doi: 10.1088/1361-6560/61/24/8908.

[37]   Z. Wang, A. C. Bovik, H. R. Sheikh, and E. P. Simoncelli, "Image quality assessment: from error visibility to structural similarity," *IEEE Transactions on Image Processing*, vol. 13, no. 4, pp. 600–612, Apr. 2004, doi: 10.1109/TIP.2003.819861.